\documentclass[12pt]{article}
\usepackage{cite,epsfig,amssymb,amsmath,graphicx,color}
\newcommand{\be}{\begin{equation}}
\newcommand{\ee}{\end{equation}}
\newcommand{\bear}{\begin{eqnarray}}
\newcommand{\eear}{\end{eqnarray}}
\newcommand{\ba}{\begin{array}}
\newcommand{\ea}{\end{array}}
\newcommand{\nn}{\nonumber}

\usepackage[normalem]{ulem}
\usepackage{varioref}

\begin{document}

\vspace{9mm}

\begin{center}
{{{\Large \bf Cosmological Perturbations of a Quartet of Scalar Fields
with a Spatially Constant Gradient}
}\\[17mm]
Seoktae Koh$^1$,~~~Seyen Kouwn$^2$,~~O-Kab Kwon$^{2,3}$, and ~~Phillial Oh$^2$\\[3mm]
{\it $^{1}$Department of Science Education, Jeju National University,
Jeju 690-756, South Korea} \\[2mm]
{\it $^{2}$Department of Physics,~BK21 Physics Research Division,
~Institute of Basic Science,\\
Sungkyunkwan University, Suwon 440-746, Korea}\\[2mm]
{\it $^{3}$Institute for the Early Universe, Ewha Womans University, Seoul 120-750, South Korea}\\[2mm]
{\tt kundol.koh@jejunu.ac.kr, ~seyen@skku.edu,~okabkwon@ewha.ac.kr,~ploh@skku.edu} }
\end{center}

\vspace{20mm}

\begin{abstract}
We consider the linear perturbations for the single scalar field
inflation model interacting with an additional triad of scalar fields.
The background solutions of the three additional scalar fields depend
on spatial coordinates with a constant gradient $\alpha$ and the ensuing evolution
preserves the homogeneity of the cosmological principle.
After we discuss the properties of background evolution including
an exact solution for the exponential-type potential,
we investigate the linear perturbations of the scalar and
tensor modes in the background of the slow-roll inflation.
In our model with small $\alpha$, the comoving  wavenumber has
{\it a lower bound}
$\sim \alpha M_{\rm P}$ to have well-defined initial quantum states.
We find that cosmological quantities, for instance, the power spectrums
and spectral indices of the comoving curvature and isocurvature perturbations,
and the running of the spectral indices have small corrections
depending on {\it the lower bound}.
Similar behaviors happen for the tensor perturbation with
the same lower bound.
\end{abstract}
\newpage

\tableofcontents

\section{Introduction}

Inflation~\cite{Guth:1980zm,Lyth:1998xn}
 has been so far very successful to account for the observational
 data~\cite{Ade:2013uln, Hinshaw:2012fq}.
It is based on the simple idea that the Universe went through a huge accelerating
expansion during the early stage of its evolution driven by a single (multiple) scalar field(s).
It provides not only spatially flat and homogeneous
universe but also  a source for small primordial perturbations which are the origin of
the large-scale structure of our Universe today.
Even though the data are compatible with the single field inflation,
there have been considerable interests in multi-field inflation.

From the point of view of cosmological perturbations~\cite{Bardeen:1980kt},
 multi-field inflation  exhibits distinctive features of
the non-Gaussianity~\cite{Maldacena:2002vr, Seery:2005gb} and generates isocurvature (entropy) perturbations in addition to adiabatic perturbations of the single scalar field model.
There exists an extensive amount of literature how these aspects are incorporated into the specific multi-field inflationary models~\cite{Polarski:1994rz,Sasaki:1995aw,Gordon:2000hv}.
In particular, the presence of isocurvature modes corresponds to relative perturbations
among the various matter fields and affects the final curvature perturbation
at the end of inflation by acting as a source term in the evolution equation
for the curvature perturbation \cite{Gordon:2000hv}.

So far, most of analysis focused on the time-dependent backgrounds in which
each of the scalar field is a function of time only. However, growing interest
in this field motivates to seek other theoretical possibilities.
One of the alternative approaches is to consider spatially dependent
backgrounds in cosmology~\cite{ArmendarizPicon:2007nr,Endlich:2012pz}.
It seems that relatively little attention has been paid to
spatially dependent backgrounds in inflation.
This might be due to the  obstacle which is inherent in these
solutions in cosmology: in general, these configurations are not
compatible with the cosmological
principle of homogeneity and isotropy.
However, the conflict can be avoided in the
nonlinear sigma models which include a triad of
scalar fields, $\sigma^a~(a=1,2,3)$.\footnote{Cosmological perturbations with
the triad of scalar fields having solutions $\sigma^a\sim x^a$
were considered before in \cite{ArmendarizPicon:2007nr,Endlich:2012pz}.
The main interest of the paper \cite{ArmendarizPicon:2007nr} was to create
statistically anisotropic and inhomogeneous perturbations
with metric perturbations. On the other hand, in \cite{Endlich:2012pz} the authors
introduced SO(3) symmetric scalar fields only without potential
for the scalar fields. The background solution for the triad of the
scalar fields is the same with ours. They considered linear perturbations and
non-Gaussianity for the model, dubbed as {\it solid inflation}.
As we see in \eqref{sig_act}, we additionally considered the inflaton
with a potential and find nontrivial results in the linear scalar perturbations.}
It is known that by correlating each scalar field
with spatial coordinate, $\sigma^a \sim  x^a$,
one can maintain the isotropy and homogeneity
in a wide class of field theory models whose Lagrangians
are  functions of
$X=-{1\over2}g^{\mu\nu}\tilde h_{ab}(\sigma)\partial_\mu \sigma^a \partial_\nu \sigma^b$
without the potential for the $\sigma^a$ fields.

The ansatz for scalar fields of a nonlinear sigma model,
$\sigma^m\sim x^m$ ($i=1,\cdots, N$)
with the extra dimension $N$,
first appeared in the higher dimensional gravity theory
and the solutions of the scalar fields trigger spontaneous compactification
of the extra dimensions~\cite{Omero:1980vx,GellMann:1984mu}.
The compactified extra space is isomorphic to the target space of  scalar fields.
A similar ansatz which break the diffeomorphism invariance was used to
give masses to gravitons as a Higgs mechanism of gravity~\cite{'tHooft:2007bf}.
Therefore, the coordinate dependent ansatz in the nonlinear sigma model in gravity theories
can be considered as a method of constructing the massive gravity
theories~\cite{Hinterbichler:2011tt}.
Recently,  the four scalar fields  were combined into de Sitter target space
and used in describing the late-time accelerating universe \cite{Lee:2009zv}.
See also \cite{Ho:2010vv,Ho:2012gp} for related topics.

Motivated by the nonlinear sigma models inherited from the higher
dimensional gravity theories, in this paper, we consider a nonlinear
sigma model with a potential as an action of matter fields
in four-dimensional gravity theory,
\begin{align}\label{sig_act}
S = \int d^4 x\sqrt{-g}\left( \frac{M_{{\rm P}}^2}{2}\, R
-\frac12 g^{\mu\nu} \tilde h_{mn}(\sigma)
\partial_\mu \sigma^m\partial_\nu \sigma^n - V(\sigma)\right),
\end{align}
where $m,\,n =1,2,3,4$ and $M_{{\rm P}}$ denotes the Planck mass,
$M_{{\rm P}} \equiv (8\pi G)^{-1/2}$.
Here $\tilde h_{mn}$ can be considered as an internal metric of a
four-dimensional Riemannian manifold.
To construct an inflationary model, we consider
a specific choice for the internal metric and potential,
\begin{align}\label{sig_res}
&\tilde h_{ab}= f(\varphi) \delta_{ab},\quad
\tilde h_{a4} =0, \quad h_{44} = 1,
\nn \\
& V(\sigma) = V(\varphi),
\end{align}
where $\varphi \equiv \sigma^4$ and $f(\varphi$) is an arbitrary
positive function of $\varphi$\footnote{It is to be pointed out that  the four fields have all positive kinetic energy unlike the previously mentioned cases of massive graviton and de Sitter target space where
the scalar field $\varphi$ has a negative kinetic energy.}.
Then the resulting model can be considered as the single scalar field
model interacting with a triad of scalar fields $\sigma^a$,
where the coupling among the inflaton $\varphi$ and the triad of scalar fields is non-minimal, $\sim f(\varphi)X$.

We try to describe the cosmological inflation in terms of the action \eqref{sig_act}.
As a background solution  with the choice of \eqref{sig_res},
we apply the ansatz $\sigma^a\sim x^a$ for the background evolution,
assuming that other dynamical fields depend on the cosmic time only.
As we discussed previously, the resulting cosmological
evolution becomes homogeneous and isotropic with the FRW metric. For the background evolution, we find an exact solution describing the power-law
inflation and try to figure out the properties of the slow-roll inflation of our model.
We consider the linearized scalar and tensor perturbations for the slow-roll inflation
with a small spatial constant gradient $\alpha$ for the triad of scalar
fields.\footnote{The three perturbed modes of the triad of
scalar fields are
decomposed into one scalar and two vector modes. }
The scalar modes have two  degrees of freedom, one from $\varphi$ and the other
from the scalar mode of the triad.  They are decoupled from
each other in the minimal coupling $(f(\varphi)=1)$ and have the form of Sasaki-Mukhanov
equation~\cite{Sasaki:1986hm} {\it in the spatially flat gauge.}
We obtain the leading contributions of $\alpha$ for
the power spectrums and spectral indices of the comoving curvature
and isocurvature perturbations.

One interesting feature of our approach is that in the slow-roll case,
the requirement of unitarity for the initial quantum state imposes {\it a lower bound}
on the comoving wave number.
The existence of this limit is a consequence of the nonvanishing
 aforementioned free parameter $\alpha$ associated with spatial condensations.

The paper is organized as follows.
In section 2, we start from the action \eqref{sig_act} with the choice \eqref{sig_res}.
Using the ansatz $\sigma^a\sim x^a$ for the triad of scalar fields,
we obtain an exact solution for the power-law inflation and  investigate the properties of  the background evolution for the slow-roll inflation.
In section 3, we write the perturbed equations for the scalar and tensor modes
and discuss the gauge conditions for the scalar perturbation.
In section 4, we describe the behaviors of the perturbations up to leading
order of $\alpha$ and obtain cosmological quantities.
We conclude in section 5.

\section{Background Evolutions}\label{sec2}

We start from the action \eqref{sig_act} with the choice \eqref{sig_res},
\begin{align}\label{action1}
S = \int d^4x \sqrt{-g}\Biggr[ {M_{{\rm P}}^2\over 2}R -{1\over2}\,
g^{\mu\nu} \partial_\mu \varphi \partial_\nu \varphi
-{1\over2} f(\varphi) g^{\mu\nu}\partial_\mu \sigma^a \partial_\nu \sigma^a -V(\varphi) \Biggr],
\end{align}
where $\sigma^a$'s have an SO(3) symmetry.\footnote{Our model is also
related to the model in \cite{Ohashi:2013mka} in which the authors introduced the
inflation scalar field with a potential and a
two-form field to explain an anisotropic inflation. This two-form field in four
dimensions is dual to a pseudo scalar field.}
This action can be considered as the  single field inflation model
with a triad of scalar fields having noncanonical kinetic terms.
We read the energy-momentum tensor as
\begin{align}\label{EMT}
T_{\mu\nu} = \partial_\mu \varphi \partial_\nu \varphi
+ f(\varphi) \partial_\mu \sigma^a \partial_\nu \sigma^a
-g_{\mu\nu}\Biggr[{1\over2}f(\varphi) g^{\alpha\beta}\partial_\alpha\sigma^a \partial_\beta\sigma^a
+{1\over2} \, g^{\alpha\beta}\partial_\alpha \varphi \partial_\beta \varphi  +V(\varphi)
\Biggr],
\end{align}
and equations of motion of the scalar fields $\varphi$ and $\sigma^a$ as
\begin{align}\label{EOM}
&\partial_\mu\partial^\mu \varphi
- {1\over2}f'(\varphi) g^{\mu\nu} \partial_\mu \sigma^a \partial_\nu \sigma^a -V_\varphi=0,
\nn \\
&{1\over\sqrt{-g}} \partial_\mu \Big( \sqrt{-g}f(\varphi) g^{\mu\nu}\partial_\nu \sigma^a \Big)=0,
\end{align}
where $V_\varphi\equiv dV/d\varphi$.

The potential in \eqref{action1} only depends on the single scalar
field $\varphi$. The reason is related with our choice
of the spatially linear background solution for the scalar fields
$\sigma^a$, which can guarantee the cosmological
principle of homogeneity with a SO(3) invariant potential. As we will see
in the next paragraph, the spatially linear solution for $\sigma^a$
in the background FRW metric
cannot be the solution of the equations of motion \eqref{EOM}
in the presence of a nonvanishing potential of $\sigma^a$.
Furthermore,  the energy-momentum tensor
becomes a function of the spatial coordinates, which means
the breakdown of the cosmological principle of homogeneity and
isotropy. Therefore, the vanishing of the potential for
$\sigma^a$ is crucial in this paper.

Now we consider the background evolution in our setup.
The background FRW metric  is given by
\begin{eqnarray}
ds^2 = -dt^2 + a(t)^2(dx^2 + dy^2 + dz^2) \,,
\end{eqnarray}
where $a(t)$ is the scale factor.
In this paper, we consider
an ansatz for the  scalar field $\sigma^a$ as
\begin{align}\label{sig_anstz}
\sigma^a =M_{\rm P}^2\alpha\, x^a,
\end{align}
where $\alpha$ is an arbitrary dimensionless constant.
Under the assumption that the scalar field $\varphi$ is spatially homogeneous, we easily see
that the ansatz \eqref{sig_anstz} satisfies the equation of motion for $\sigma^a$
in \eqref{EOM}.
Remaining equations of motion for
$g_{\mu\nu}$ and $\varphi$ are given by
\begin{align}
&H^2 = \frac{\rho}{3M_{\rm P}^2}  \,,\nonumber  \\
&\dot{H} = -{\rho + p\over 2M_{\rm P}^2}  \,, \label{back_EOM}  \\
&\ddot{\varphi} +3 H\dot{\varphi} +{3M_{\rm p}^4\alpha^2 \over 2a^2}f_\varphi
+ V_\varphi= 0 \,, \nonumber
\end{align}
where $f_\varphi\equiv df/d\varphi$, $H\equiv \dot{a}/a$, and
\begin{align}\label{rhop}
\rho &= {1\over 2}\dot{\varphi}^2 + { 3 M_{\rm p}^4
\alpha^2 \over 2a^2 }f + V,
\nn \\
p &= \frac12\dot\varphi^2 -{ M_{\rm p}^4 \alpha^2 \over 2a^2 }f - V.
\end{align}
As we see from these background equations, the homogeneity of the background evolution
is not spoiled though we take the spatially dependent ansatz for $\sigma^a$ in \eqref{sig_anstz}.
The $f$-dependent terms in \eqref{rhop} are originated from the contributions of $\sigma^a$. Therefore, the equation of state of $\sigma$-fields
is $w_\sigma = -1/3$. For concreteness, we fix the function $f(\varphi)$ as
\begin{align}\label{fix-f}
f(\varphi) = e^{2\xi\varphi/M_{{\rm p}}},
\end{align}
where $\xi$ is an arbitrary constant.
In the case $\xi=0$, i.e., $f(\varphi)=1$, the role
of the $\alpha^2$-dependent term in \eqref{back_EOM} is the
same as the curvature constant determining the spatial curvature  in the
Friedmann equations. Since we are considering the positive
$\alpha^2$, the background evolution under the spatial dependent
solution \eqref{sig_anstz} corresponds to that of the open universe.
However, our model is different from the single field inflation model
on the background metric of the open universe.
As we will see later, there are nontrivial roles of $\sigma^a$ in
the perturbation level.

The main purpose of this work is to investigate the
behaviors of our model \eqref{action1} from the point of view of
 linear perturbations.
Before we move on to the subject of the perturbation,
we describe the background evolutions of our model
by considering two inflationary scenarios,
the power-law inflation and the slow-roll inflation.

\subsection{Power-law inflation}\label{plinf}

It is well-known that under an exponential-type potential the single scalar field model
has an exact solution describing the power-law inflation~\cite{Lucchin:1984yf}.
Comparing with the single scalar field model,
there are $\alpha$-dependent terms in \eqref{back_EOM}, which are
originated from the spatially dependent background solution for $\sigma^a$.
As we will see in this subsection, the exact solution for the power-law inflation
in the single scalar field model can be a solution of the equations of motion in \eqref{back_EOM} with some deformations by parameters, $\xi$ and $\alpha$.

 As in the case of the single scalar field model, we consider an exponential-type potential
\begin{align}\label{exp_pot}
V(\varphi)=V_0e^{-\lambda \varphi/M_{{\rm P}}},
\end{align}
where $V_0$ and $\lambda$ are arbitrary constants.
We consider an ansatz which describes the power-law inflation,
\begin{align}\label{power_sol}
a(t)=a_0 (M_{{\rm P}}\, t)^n,\qquad \varphi(t)=M_{{\rm P}}\left(\frac{2}{\lambda} \ln (M_{{\rm P}} \,t )
+ \varphi_0 \right),
\end{align}
where $a_0$, $n$, and $\varphi_0$ are dimensionless constants.

When $\alpha=0$, the background equations
reduce to those of the single scalar field model.
The constant parameters in the power-law solution \eqref{power_sol}
are determined as
\begin{align}\label{alphaz}
n=\frac2{\lambda^2},\qquad
\varphi_0 =\frac1{\lambda}\ln\left(
\frac{V_0\lambda^4}{2M_{{\rm P}}^4(6-\lambda^2)}\right),
\end{align}
and $a_0$ becomes a free parameter.

However, in the case of $\alpha\ne 0$,
all constant parameters of the power-law solution \eqref{power_sol} are fixed by
\begin{align}\label{alphanz}
n&=1+{2\xi\over \lambda},
\nn \\
a_0^2 &= \frac{\alpha^2\lambda^2}{2(\lambda^2+2\lambda\xi -2)}
\left(\frac{\lambda^2V_0}{2M_{{\rm P}}^4(6\xi^2 + 3\lambda\xi +2)}\right)^{2\xi/\lambda},
 \\
\varphi_0 &= \frac1{\lambda}\ln \left(\frac{\lambda^2V_0}{2M_{{\rm P}}^4(6\xi^2
+ 3\lambda\xi +2)}\right).\nn
\end{align}
The  positive definiteness of the right-hand side of the second condition in
\eqref{alphanz} restricts the value of $\lambda$,
\begin{align}\label{lam_xi}
\lambda >- \xi + \sqrt{\xi^2 +2}\quad {\rm or}\quad \lambda < -\xi - \sqrt{\xi^2 +2}.
\end{align}
In order to have an accelerating unverse, we also have a   constraint, $\frac{\xi}{\lambda}>0$, from the first
relation in \eqref{alphanz}.
Therefore, there are two possibilities for the parameters,
$\lambda$ and $\xi$, i.e., $\lambda>0$ $\&$ $\xi>0$ or
$\lambda<0$ $\&$ $\xi<0$. In the case $\lambda>0$ $\&$ $\xi>0$ we have the first relation
in \eqref{lam_xi},
while in the $\lambda<0$ $\&$ $\xi<0$ case we have the second one
in \eqref{lam_xi}. Then the range of $n$ for both cases can be expressed in terms of $\xi$ as follows,
\begin{align}
1<n<1+\xi^2 + \sqrt{\xi^4 +2\xi^2} \,.
\end{align}
In the case $\xi=0$, $n$ is fixed to unity.
Therefore, there is no inflation in this case.
In this paper, we consider $\lambda>0$ $\&$ $\xi>0$ case.

The power-law solutions in \eqref{alphaz} and \eqref{alphanz}
are special solutions of the background equations of \eqref{back_EOM} since we started from a special ansatz in obtaining the solutions.
For this reason, it is understandable that one cannot obtain the solution \eqref{alphaz} by taking $\alpha\to 0$
in \eqref{alphanz}. That is, these two solutions satisfy different initial conditions
in the $\alpha\to 0$ limit.
The general solution under the exponential-type potential \eqref{exp_pot} for the
single scalar field model was obtained in \cite{Andrianov:2011fg}.

\subsection{Slow-roll inflation} \label{bgsr}

Next, we consider the slow-roll inflation with  a potential
satisfying the slow-roll approximations,
\begin{align} \label{sr_approx}
\dot{\varphi}^2/2 \ll V,
\qquad \ddot{\varphi} \ll 3H\dot{\varphi}.
\end{align}
To reflect these approximations, we introduce the slow-roll parameters,
\begin{align} \label{srparam}
\epsilon \equiv{\dot{\varphi}^2\over 2 M_{\rm P}^2 H^2}, \qquad
\eta \equiv{V_{\varphi\varphi}\over 3H^2}.
\end{align}
With the approximations in (\ref{sr_approx}),
the background equations of motion of  (\ref{back_EOM}) reduce to
\begin{align}\label{sr-bgeq2}
&  H^2 \simeq{\alpha^2 M_{\rm P}^2 \over 2 a^2} f+  \frac{1}{3M_{{\rm P}}^2}V, \nn \\
&  3H\dot{\varphi} + \frac{3 \alpha^2 M_{\rm P}^4}{2a^2} f_{\varphi}
+ V_{\varphi} \simeq 0,
\end{align}
where $f_\varphi = \frac{2\xi}{M_{{\rm P}}}
e^{2\xi\varphi/M_{{\rm P}}}$ for the case \eqref{fix-f}.
The background dynamics of $\varphi$ is governed by
the two papameters $\alpha$ and $\xi$ as well as the shape of
potential $V(\varphi)$.
\begin{figure}
\centering
\includegraphics[width=0.5\textwidth]{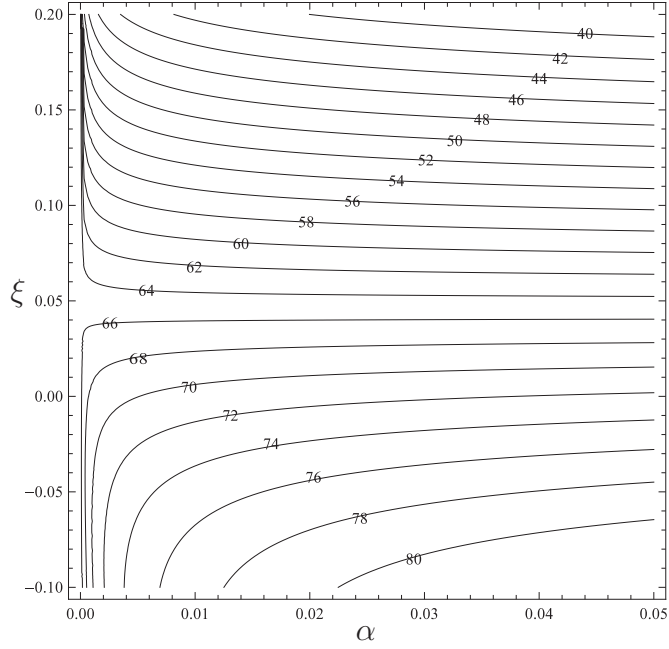}
\caption{Plots of $e$-foldings for the quadratic potential
$V= \frac12 m^2\varphi^2$ on the parameter space of
$\alpha$ and $\xi$.
We choose the initial condition as $a_0=1$,
$\varphi_0=16M_{\rm P}$, and $\dot{\varphi}_0=0$
with $m=10^{-6}M_{\rm P}$.}
\end{figure}

We can compute the number of $e$-foldings using the slow-roll equations
of motion of \eqref{sr-bgeq2},
\begin{align}\label{efdN}
\mathcal{N} = \int_{t}^{t_e} Hdt = -\int_{\varphi_e}^{\varphi_i}
 \frac{H}{\dot{\varphi}} d\varphi
\simeq \frac1{M_{{\rm P}}^2} \int_{\varphi_e}^{\varphi_i}
\frac{\frac{V}{3}+\frac{\alpha^2 M_{{\rm P}}^4 f}{2 a^2}  }{
\frac{V_\varphi}{3}+\frac{\alpha^2 \xi M_{{\rm P}}^3 f}{ a^2}  }\,
d\varphi,
\end{align}
where $\varphi_i$ and  $\varphi_e$ are
the  values of $\varphi$ at the initial time and
at the end of inflation, respectively.
For a fixed value of $\alpha$, the number of $e$-foldings
decreases with  increasing $\xi$ since  the $\xi$-dependent
term only appears in the denominator in \eqref{efdN}.
For concreteness, we consider a massive scalar field with a
potential
\begin{align}\label{quadV}
V(\varphi) = \frac12 m^2 \varphi^2.
\end{align}
Then the number of $e$-foldings in \eqref{efdN} is given by
\begin{align}\label{efdN2}
\mathcal{N} \simeq \frac1{M_{{\rm P}}^2} \int_{\varphi_e}^{\varphi_i}
\frac{\frac{\varphi}{2}\left(1+\frac{\alpha^2 M_{{\rm P}}^4f}{
a^2 m^2 \varphi^{2}}\right)}{1+ \frac{\alpha^2\xi
M_{{\rm P}}^3f}{ a^2 m^2 \varphi}}\, d\varphi.
\end{align}
For a small value of $\alpha$, we can expand the number of
$e$-foldings as
\begin{align}\label{efdN3}
\mathcal{N} \simeq \frac1{2M_{{\rm P}}^2} \int_{\varphi_e}^{\varphi_i}
\varphi\Big[1-\left(\xi - \frac{M_{{\rm P}}}{\varphi}\right)\,
\frac{\alpha^2 M_{{\rm P}}^3f}{ a^2 m^2 \varphi}+
{\cal O}\left(\alpha^4\right)\Big]d\varphi .
\end{align}
In this limit, the $\alpha$-dependent term in \eqref{efdN3} is
also small and the contribution of this term to the number of
$e$-foldings is negligible and independent of the magnitude of
$\xi$. However, the sign of $\xi - M_{{\rm P}}/\varphi$
determines the negative or positive contributions to the number
of $e$-foldings. Therefore, there is a critical point
for the value of $\xi$ near
\begin{align}\label{cxi}
\xi \approx M_{{\rm P}}/\varphi.
\end{align}
The scalar field $\varphi$ is varying as the cosmic time evolves. However,
it is almost a constant at the early stage of the inflation, according to
our assumption of the slow-roll inflation.
For this reason, the value of $\xi$ in \eqref{cxi} can be the critical
point. The behaviors of the $e$-foldings for the quadratic potential
\eqref{quadV} are given graphically in Fig.1.

One of the well-known properties of the slow-roll inflation
($\alpha=0$) is that there exists a late time attractor behavior,
such that the time evolution of the inflaton scalar field becomes independent of
the initial conditions. In our case ($\alpha \ne 0$) in \eqref{back_EOM},
the time evolution behavior at the early time is different from that of
the single field case due to some contributions of the $\alpha$-dependent terms
in \eqref{back_EOM}. As time goes on, however, the contributions become smaller
quickly since those terms are proportional to $1/a^2$. Therefore, we have similar
attractor behavior in our case as well.

\section{Linear Perturbations}

In this section, we  discuss the linear perturbations
for our model.
We introduce a new gauge invariant variable in the scalar perturbation, which is originated
from the space dependent background solutions of the triad of scalar fields, $\sigma^a$.
After we take {\it the spatially flat gauge} which is convenient
in our later analysis for the linear perturbation,
we write the general form of the comoving curvature perturbation
and the isocurvature perturbation.  We also discuss the perturbation for the tensor mode.

\subsection{Scalar mode}

In this subsection we consider the linear scalar perturbation for our model \eqref{action1}.
The linear scalar perturbation of the metric is given by
\begin{align}\label{met_pert}
ds^2 = -\left(1+2A\right)dt^2 +2a\, \partial_iB\,dt dx^i
+a^2\Big[\left(1-2\psi\right)\delta_{ij}+2\partial_i\partial_jE \Big]dx^i dx^j,
\end{align}
where $A$, $B$, $\psi$, and $E$ are four scalar perturbation modes.
We also consider perturbations of the scalar fields,
\begin{align}
\varphi(t,x) &= \varphi(t) + \delta\varphi(t,x) \,,
\nn \\
\sigma^a(t,x) &= \sigma^a(x) + \delta\sigma^a(t,x) \,.\label{phisig}
\end{align}

Before we write the perturbation equations for scalar modes,
we discuss the contribution of $\delta \sigma^i$ to scalar modes.
At first, we read the $(0,i)$-component of the perturbed Einstein equation
$\delta G^0_{~i} = M_{\rm P}^{-2} \,\delta T^0_{~i}$ as
\begin{align}\label{G0i}
-2\,\partial_i\left(HA+\dot{\psi}\right)
= M_{\rm P}^{-2}\delta T^0_{~i}
= M_{\rm P}^{-2}\left(-\partial_i(\dot\varphi\delta\varphi)
-\alpha f \delta\dot\sigma^i + \frac{\alpha^2f}{a}\partial_i B\right).
\end{align}
Taking the curl of both sides of \eqref{G0i}, we obtain
\begin{align}
\epsilon_{ijk}\partial^j \delta\sigma^k=0.
\end{align}
From this relation, we see that $\delta\sigma^i$ appearing in the linear scalar
perturbation equations has no contribution to the perpendicular mode.
For this reason, we can set\footnote{In general,
$\delta\sigma^i$ is decomposed as
$\delta\sigma^i = \frac1{k}\partial_i u + \delta\sigma^i_\perp$
with the perpendicular mode
$\delta\sigma^i_\perp$ satisfying $\partial_i\delta\sigma_{\perp}^i=0$
in the linear perturbation. When we consider the perturbation of the
vector mode, $\delta\sigma_\perp^i$ can contribute to the
perturbed equations. It is well-known that after the inflation
the universe enters into the matter-dominated era, and vector modes
should decay leaving no detectable imprints.
As we discussed in section 2, the background in our model
has similar attractor behavior with that of the single field model.
For this reason, the vector modes in our model can decay as well.
We do not consider the perturbation
of the vector mode in this paper.}
\begin{align}\label{partu}
\delta\sigma^i = \frac1{k}\partial_i u,
\end{align}
where $k$ is the comoving wave number and $u$ is the perturbed
scalar mode.  Here we introduce $1/k$ factor
in order to define the canonical kinetic term for the scalar mode $u$
in the perturbed Lagrangian.
Then we can write the variation of the $(0,i)$-component of the energy
momentum tensor as
\begin{align}
\delta T^0_{~i} = \partial_i \delta q,
\end{align}
where the scalar part of the three momentum $\delta q$ is given by
\begin{align}\label{defdq}
\delta q &= -\dot\varphi \delta\varphi
-{M_{\rm p}^2\alpha\over k}f\dot{u} +{M_{\rm p}^4\alpha^2 \over a} f  B.
\end{align}

\subsubsection{Perturbation equations}

Inserting the relations for perturbed modes, \eqref{met_pert}, \eqref{phisig}, and \eqref{partu} into
the Einstein equation, we obtain the linearized perturbed Einstein
equations $\delta G^\mu_{~\nu} = \frac1{M_{{\rm P}}^2}\delta T^\mu_{~\nu}$:
\begin{align}\label{Eeq}
(0,0):\,&3H\left(HA+\dot\psi\right)
-\frac{1}{a^2}\nabla^2\left[\psi + H\Big(a^2\dot E- aB\Big)\right]
=-{\delta\rho\over 2M_{\rm p}^2} \,,
\nonumber \\
(0,i):\,&  HA+\dot\psi
=-{\delta q\over 2M_{\rm p}^2}  \,,
\nonumber \\
(i,j):\,&2\Biggr[\,\ddot{\psi}
+3H\Big(HA+\dot{\psi}\Big)+H\dot{A}+2\dot{H}A
+{1\over 2a^2}\nabla^2D \Biggr]\delta^i_{~j}
-{1\over a^2}\partial_i\partial_jD \nonumber \\
&={1\over M_{\rm P}^2}\biggr(
\delta p \,\delta^i_{~j} +\Pi^i_{~j}
\biggr)
 \,,
\end{align}
where
\begin{align}\label{drdp}
\delta\rho &=  \dot\varphi (\dot{\delta\varphi}-\dot\varphi A)
+{\partial V\over \partial \varphi}\delta\varphi
+{M_{\rm p}^4\alpha^2\over a^2}f
\Biggr(\,\nabla^2 \Big({u\over \alpha k M_{\rm p}^2}-E\Big)+3\psi
+{3\xi\delta\varphi\over M_{\rm p}}\Biggr),
\nn \\
\delta p &= \dot{\varphi}\delta\dot{\varphi}- A \dot{\varphi}^2
-V_\varphi\delta\varphi -{M_{\rm p}^4\alpha^2\over 3a^2}
f\left( \nabla^2
\Big({u\over \alpha k M_{\rm p}^2}-E\Big)  +3\psi
+{3\xi\delta\varphi\over M_{\rm p}}\right),
\nn \\
D &= A-\psi -H\Big(a^2\dot{E}-aB\Big)
-{d\over dt}\Big(a^2\dot{E}-aB\Big) \,, \nn \\
\Pi^i_{~j} &= { 2M_{\rm P}^4\alpha^2\over a^2}f
\Biggr[
 \partial_i\partial_j \Big({u\over \alpha k M_{\rm P}^2}-E\Big)
 -{1\over3}\nabla^2 \Big({u\over \alpha k M_{\rm P}^2}-E\Big)
 \Biggr]
 \,.
\end{align}

From \eqref{EMT} and \eqref{EOM}, we obtain the perturbed equations for $\delta\varphi$ and $u$
without gauge fixing,
\begin{align}\label{phiu}
& \delta\ddot{\varphi}+3H\delta\dot{\varphi}-{1\over a^2}\nabla^2 \delta\varphi
+{2M_{\rm p}^2\alpha^2\,\xi \over a^2}f
\Big({\nabla^2u\over \alpha k M_{\rm p}}+3\xi\delta\varphi\Big)
+{\partial^2V\over\partial\varphi^2}\delta\varphi \nonumber \\
&~~ = 2A\ddot{\varphi}+6HA\dot{\varphi}+\dot{\varphi}
\Big[\dot{A}+3\dot{\psi}+\nabla^2\Big({B \over a}-\dot{E}\Big)\Big]
-{2M_{\rm P}^3\alpha^2\,\xi \over a^2}f\Big(3\psi-\nabla^2E\Big),
\nn \\
&\ddot{u} + \Biggr(3H+{2\xi \dot{\varphi}\over M_{\rm P}}\Biggr) \dot{u}- {1\over a^2}\nabla^2 u
={\alpha k M_{\rm P}^2\over a^2}\Biggr[ a \dot{B}  + 2a\Biggr(H+{\xi\dot{\varphi}\over M_{\rm P}}\Biggr)B + {2\xi\delta\varphi\over M_{\rm P}}
+  \big(A-\psi-\nabla^2 E\big) \Biggr].
\end{align}

Using the relations in \eqref{defdq} and \eqref{drdp}, we can
write the comoving curvature perturbation~\cite{Lukash:1980iv,Lyth:1984gv}
and the isocurvature perturbation
for multiple scalar fields~\cite{Gordon:2000hv} as
\begin{align}\label{RS}
{\cal R} \equiv \psi -\frac{H}{\rho + p}\delta q,
\qquad
{\cal S} \equiv H\left(\frac{\delta p}{\dot p}
- \frac{\delta\rho}{\dot\rho}\right).
\end{align}

\subsubsection{Gauge condition}

As we see in \eqref{partu}, we have one additional  scalar degree of
freedom due to the contribution of $\delta\sigma^a$.
Therefore, we have six degrees of freedom for the scalar perturbation modes.
Eliminating the gauge degrees of freedom and imposing the constraints
of the Einstein equation, we have two physical degrees of freedom.
Since we have considered the spatially dependent background
solution for $\sigma^a$, we have slightly different gauge
invariant quantities from the well-known multi-field models.

Under the change of coordinate,
\begin{align}
x^{\mu}\,\,\rightarrow \,\, x^\mu + \beta^\mu, \quad
(\beta^0 =\beta,\,\beta^i =\delta^{ij}\partial_j\gamma),
\end{align}
the scalar modes of the perturbed metric \eqref{met_pert} transform as
\begin{align}\label{str1}
&A\,\,\rightarrow \,\, A-\dot\beta,
\nn \\
&B \,\,\rightarrow \,\, B + \frac{\beta}{a} - a\dot\gamma,
\nn \\
&\psi \,\,\rightarrow \,\, \psi - H\beta,
\nn \\
&E\,\,\rightarrow \,\, E-\gamma.
\end{align}
The scalar modes of matter fields also transform as
\begin{align}\label{str2}
\delta\varphi  \,\,&\rightarrow \,\, \delta\varphi -\beta^\mu\partial_\mu\varphi
=\delta\varphi -\dot\varphi\beta ,
\nn \\
\delta\sigma^a\,\,&\rightarrow \,\, \delta\sigma^a - \beta^\mu\partial_\mu\sigma^a
= \frac1{k}\delta^{ab}\partial_b u -\alpha M_{{\rm P}}^2\delta^{ab}\partial_b\gamma,
\end{align}
where we used the background solution
$\sigma^a = \alpha M_{{\rm P}}^2 x^a$ of \eqref{sig_anstz}
in the last step.
From \eqref{str2} we can read off the gauge transformation of the scalar mode
$u$ as
\begin{align}\label{str3}
u \,\,\rightarrow \,\, u -k\alpha M_{{\rm P}}^2\gamma.
\end{align}
From these transformation rules, we can obtain several gauge
invariant quantities.
For later convenience we define those quantities,
\begin{align}
\Phi &\equiv A + \frac{d}{dt}\left( a(B- a\dot E)\right),
\nn \\
\Psi &\equiv \psi + a H( B- a\dot E),
\nn \\
Q_{\varphi} &\equiv \delta\varphi - \frac{\dot\varphi}{H}\psi,
\nn \\
Q_u &\equiv u- \alpha k M_{{\rm P}}^2E.
\end{align}
The new quantity $Q_u$ was introduced due to the spatial dependence of the
background field $\sigma^a$.

To solve the perturbed equations \eqref{Eeq} and \eqref{phisig},
we take {\it the spatially flat gauge} $(\psi=0\,\,\&\,\, E=0)$.
In this gauge the perturbed equations are reduced to
\begin{align}\label{sflat}
&\ddot{Q}_\varphi+3H\dot{Q}_\varphi
+\Biggr( {k^2\over a^2}+{\dot{\varphi}V_\varphi\over M_{\rm P}^2 H}+V_{\varphi\varphi}
+{6M_{\rm P}^2\alpha^2 \xi^2\over a^2} f
+{2M_{\rm P} \alpha^2 \xi \dot{\varphi} \over a^2 H} f   \Biggr)Q_{\varphi}
\nonumber \\
&\hskip 3cm - \frac{2\xi \alpha k M_{{\rm P}}}{a^2}f Q_u
+2\Biggr({\dot{H}\dot{\varphi}\over H}-\ddot{\varphi}\Biggr)A=0,
\nn \\
&\ddot{Q}_{u}+\Biggr(3H+{2\xi\dot{\varphi}\over M_{{\rm P}}}\Biggr)\dot{Q}_u
+\Biggr({k^2\over a^2}
+ {2\alpha^2 M_{\rm p}^2 f\over a^2}\Biggr)Q_u
- \frac{2\xi \alpha k M_{{\rm P}}}{a^2}\Big(Q_\varphi+aB\dot{\varphi}\Big)=0,
\end{align}
where  $Q_\varphi = \delta\varphi$, $Q_u = u$.
The scalar modes and $A$ and $B$
satisfy the constraints which are given from the first two
equations of \eqref{Eeq},
\begin{align}\label{cAB}
&3AH^2-{k^2BH\over a}=\frac1{2M_{\rm p}^2}\Big(A\dot{\varphi}^2-\dot{\varphi}\dot{Q}_\varphi
- V_\varphi Q_\varphi\Big) +\Big(\alpha k Q_u-3\alpha^2\xi M_{\rm p}Q_\varphi\Big)
{f\over 2a^2},
\nn \\
&2AH={\dot{\varphi} Q_\varphi\over M_{\rm p}^2}
+\Biggr(\frac{\alpha}{k}Q_u-{\alpha^2 M_{\rm p}^2 B\over a }
\Biggr)f.
\end{align}

The comoving curvature and isocurvature perturbations
of  \eqref{RS} are written {\it in the spatially flat gauge} as
\begin{align}
{\cal R}&=H\Biggr[{
\dot{\varphi}Q_\varphi
-\alpha M_{\rm P}^2f
\big({\alpha M_{\rm P}^2 B\over a}-{{\dot  Q_u }\over k}\big)
\over {\dot\varphi}^2
+{\alpha^2M_{\rm P}^4\over a^2}f
} \Biggr],
\nn \\
{\cal S} &=H\Biggr[
{
\dot{\varphi}(\dot{Q}_\varphi-A\dot{\varphi})-V_\varphi Q_\varphi
+{\alpha^2 M_{\rm P}^4\over 3a^2}f\Big({k Q_u\over a M_{\rm P}^2}-{3\xi Q_\varphi\over M_{\rm P}}\Big)
\over
\dot{\varphi}(\ddot{\varphi}-V_\varphi)
+{\alpha^2M_{\rm P}^4\over a^2}f\Big(H- {\xi\dot{\varphi}\over M_{\rm P}}\Big)
}
-
{
\dot{\varphi}(\dot{Q}_\varphi-A\dot{\varphi})+V_\varphi Q_\varphi
-{\alpha^2 M_{\rm P}^4\over a^2}f\Big({k Q_u\over a M_{\rm P}^2}-{3\xi Q_\varphi\over M_{\rm P}}\Big)
\over
\dot{\varphi}(\ddot{\varphi}+V_\varphi)
-{3\alpha^2M_{\rm P}^4\over a^2}f\Big(H-{\xi\dot{\varphi}\over M_{\rm P}}\Big)
}
\Biggr].\label{ICP}
\end{align}
In the case $\alpha=0$, the quantities ${\cal R}$ and ${\cal S}$ are
reduced to those of the single scalar field~\cite{Gordon:2000hv}.

\subsection{Tensor mode}

In this subsection, we investigate the linear perturbation of the tensor mode for
our model on the background described by the equations of  \eqref{sig_anstz} and \eqref{back_EOM}.
Since the tensor mode is decoupled from the scalar and vector ones in the linear perturbation
theory, we consider the perturbed metric in the
conformal time to describe the tensor perturbation,
\begin{align}\label{tmetric}
ds^2 = a^2(\tau)\left( -d\tau^2 + (\delta_{ij} + h_{ij})dx^idx^j\right),
\end{align}
where $dt = ad\tau$.
The tensor modes $h_{ij}$ satisfy the following conditions,
\begin{align}\label{tconst}
\partial_i h_{ij} = 0, \qquad h_{ii} = 0.
\end{align}
Since the tensor modes $h_{ij}$ have two degrees of freedom, which are identified with
the two polarizations of gravitational waves, we expand the tensor modes as
\begin{align}\label{tpol}
h_{ij}(\tau,\vec x) = \frac2{M_{{\rm P}}}\sum_{\lambda=+,-}
\frac{\tilde\mu_\lambda(\tau,\vec x)}a \epsilon_{ij}^\lambda,
\end{align}
where $\epsilon_{ij}^\lambda$ is the polarization tensor satisfying the orthogonality condition
$\epsilon_{ij}^\lambda\epsilon^{ij}_{\lambda'}= \delta^\lambda_{\lambda'}$.
Inserting \eqref{tmetric} and \eqref{tpol} into the action \eqref{action1} on our background
and solving the equation of the motion of $\tilde\mu_\lambda$, we see
that the Fourier mode of $\tilde\mu_\lambda$ satisfies
\begin{align}\label{tmas}
\mu_\lambda''(\tau;k) + \left(k^2 - \frac{a''}{a}\right)
\mu_\lambda(\tau;k) = 0,
\end{align}
where $'\equiv d/d\tau$ and $\mu_\lambda$ satisfies the normalization condition,
\begin{align}
\mu_\lambda{\mu'}^{*}_\lambda - \mu_\lambda^*{\mu'}_\lambda=i.
\end{align}
Using the relations for the background evolution in \eqref{back_EOM},
we obtain a general differential equation for the tensor mode for our case,
\begin{align}
\mu''_\lambda + \Big(k^2 - \frac12 M_{{\rm P}}^2 \alpha^2 f + \frac1{6M_{{\rm P}}^2}
\big({\varphi'}^2 - 4 a^2 V\big)\Big)\mu_\lambda =0.
\end{align}

\section{Perturbations in the Slow Roll Inflation}

As we see from the perturbed equations \eqref{sflat} and \eqref{cAB}
in the spatially flat gauge,
the perturbed modes $Q_\varphi$ and $Q_u$ are completely decoupled
in the $\alpha \to 0$ limit.
Moreover, the resulting curvature perturbations,
${\cal R}$ and ${\cal S}$ in \eqref{ICP},
are independent of $Q_u$.
Therefore, in the $\alpha \to 0$ limit, the linear perturbation of our model
is the same as that of the single scalar field model.
For this reason, in this paper,
we only consider the case with nonvanishing $\alpha$.

When we consider the linear perturbations with nonvanishing $\xi$
and $\alpha$, in general the modes $Q_\varphi$ and
$Q_u$ are not decoupled.
However, in this coupled case it is very difficult to obtain some
cosmological quantities, such as power spectrums
and scalar/tensor spectral indices.
We leave this general case for further investigation.
In this paper, we restrict the linear perturbations to the case of
the slow-roll inflation at a small value of $\alpha$
with vanishing $\xi$.
As we will see in this case, the
modes $Q_\varphi$ and $Q_u$ are
decoupled. However, ${\cal R}$ and ${\cal S}$ still have dependence on both  $Q_\varphi$ and $Q_u$.

In this paper, we will not consider the perturbation for the
power-law inflation. The reason is the following.
For the case of $\alpha \ne 0$ with $\xi=0$,
the background equations in \eqref{back_EOM}
have a special solution given in \eqref{power_sol}
and \eqref{alphanz} for the exponential-type potential
\eqref{exp_pot}.
In this case, the solutions are reduced to
\begin{align}
\varphi=\frac{2M_{{\rm P}}}{\lambda} \ln \left(\frac{\sqrt{V_0}\lambda t}{2 M_{\rm P}}\right),
\qquad a(t)=\frac{M_{\rm P} \alpha\lambda}{\sqrt{2\lambda^2 -4}}\,t \,.
\end{align}
The scale factor is a linear function of the cosmic time $t$,
and so accelerating universe is not allowed.
Presumably, if we can obtain the general solution like the
work for the single scalar with exponential-type potential
in \cite{Andrianov:2011fg}, we may have an inflation era even
in the case $\xi=0$. However, it is beyond the range of this paper.
Due to this reason, we do not consider the cosmological
perturbation for the power-law inflation.
In this section, we concentrate on the perturbation
for the slow-roll inflation described in section 2.

\subsection{Scalar mode}

In the slow-roll approximation at a small value
of $\alpha$ with $\xi=0$, the perturbed equation \eqref{sflat}
are reduced to
\begin{align}\label{QpQu}
&\ddot{Q}_\varphi+3H\dot{Q}_\varphi
+\Biggr( {k^2\over a^2}+{\dot{\varphi}V_\varphi\over M_{\rm P}^2 H}+V_{\varphi\varphi}
   \Biggr)Q_\varphi\simeq0,
\nn \\
&\ddot{Q}_u+3H\dot{Q}_u
+\Biggr({k^2\over a^2} + {2\alpha^2 M_{\rm p}^2\over a^2} \Biggr)Q_u
\simeq 0,
\end{align}
where we keep the leading order in slow-roll parameters of   (\ref{srparam}).
Changing the cosmic time $t$ into  the conformal time $\tau$ and
introducing the Sasaki-Mukhanov variables,
\begin{align}\label{VUQ}
{\cal V} = a Q_\varphi \,,~~~ {\cal U} = a Q_u \,,
\end{align}
we obtain the Sasaki-Mukhanov equations from
\eqref{QpQu},
\begin{align}\label{VU}
&{\cal V}_k''+ \Biggr(\,k_1^2 -{\mu_1^2
-\frac14\over \tau^2} \Biggr){\cal V}_k =0 ,\nn\\
&{\cal U}_k''+ \Biggr(\,k_2^2 -{\mu_2^2
-\frac14 \over \tau^2} \Biggr){\cal U}_k =0.
\end{align}
The parameters in \eqref{VU} are given by
\begin{align}\label{wavevec}
k_1^2 &\equiv k^2-{\alpha^2 M_{{\rm P}}^2\over6},
\nn \\
k_2^2 &\equiv k^2+{11\alpha^2 M_{{\rm P}}^2\over6},
\nn \\
\mu_1 &\simeq {3\over2}+3\epsilon-\eta,
\nn \\
\mu_2 &\simeq {3\over2}+\epsilon.
\end{align}
The $k_1$ and $k_2$ can be considered as the effective wavevectors
for the ${\cal V}_k$ and ${\cal U}_k$ modes, respectively.
In driving the Sasaki-Mukhanov equations in \eqref{VU}, we used the background
equations \eqref{sr-bgeq2} in the slow-roll approximation
of the subsection \ref{bgsr}. The conformal time $\tau$ has the following relation
\begin{align}\label{tauH}
\tau = \int \frac{dt}{a} = -\frac{1}{aH} + \int\frac{\epsilon^H}{a^2 H}da
\simeq -\frac1{aH}\left(1+\epsilon
+ \frac{\alpha^2 M_{{\rm P}}^2}{6 a^2H^2}\right),
\end{align}
where  the relation
\begin{align}\label{epsH}
\epsilon^H= -\frac{\dot H}{H^2}\simeq \epsilon
+ \frac{\alpha^2 M_{{\rm P}}^2}{2a^2 H^2}\,,
\end{align}
was used.  Here we keep the leading order in the slow-roll parameters,
$\epsilon$ and $\eta$, and  a small gradient constant $\alpha$.
We will discuss the range of $\alpha$ in our approximation later.
Then, from \eqref{tauH}, we obtain
\begin{align} \label{ch_time}
{\cal H}^2 &\simeq
{1+2\epsilon \over\tau^2} + \frac{\alpha^2M_{{\rm P}}^2}{3},
\end{align}
where ${\cal H}=a'/a$.

Assuming that the slow-roll parameters are constants,
we have the exact solutions for the equations of \eqref{VU},
 \begin{align}\label{VUsol}
 {\cal V}_k &=  \sqrt{-\tau} \biggl[c_{1k} H^{(1)}_{\mu_1}
 (-k_1\tau)
 + d_{1k} H^{(2)}_{\mu_1} (-k_1\tau) \biggr] e_1(k),
 \nn  \\
 {\cal U}_k &=  \sqrt{-\tau} \biggl[ c_{2k} H^{(1)}_{\mu_1}
 (-k_2\tau)
 + d_{2k} H^{(2)}_{\mu_2} (-k_2\tau) \biggl] e_2(k),
  \end{align}
where $H_{\mu}^{(i)}(z)$ ($i=1,2$) are  the first and second kinds of the Hankel functions,
and $e_i(k)$'s are independent Gaussian random variables satisfying
\begin{align}\label{ranGau}
\langle e_i(k)\rangle =0, \qquad
\langle e_i(k) e_j(k')\rangle = \delta_{ij} \delta(k- k').
\end{align}
Here, the angled brackets denote ensemble averages.

In order to determine the integration constants
$c_{ik}$ and $d_{ik}$, we have to fix the initial vacuum state
at $k\tau \rightarrow -\infty$.
As we see in \eqref{VU} and \eqref{VUsol},
the two differential equations and the corresponding solutions
have the same forms. Due to the overall normalization of $u$
mode given in \eqref{partu}, ${\cal V}_k$ and ${\cal U}_k$ modes
have the same normalizations.
Therefore, we can concentrate on the solution of ${\cal V}_k$ only
and apply the result to the case of ${\cal U}_k$.

The mode for ${\cal V}_k \equiv \tilde v_k e_1(k) $ satisfies the normalization condition,
$\tilde v_k\tilde v_k^{*'} - \tilde v_k^* \tilde v_k^{'} = i$
which is originated from the quantization condition of the mode.
Due to this normalization, we have the relation between the
integration constants,
\begin{align}\label{norcon}
|c_{1k}|^2 - |d_{1k}|^2 = \frac{\pi}{4}.
\end{align}
We adopt the Bunch-Davies vacuum for the initial perturbation mode
at $\tau\ll 0$ by taking the positive energy mode.
Then the initial mode is given by
\begin{align}\label{inicon}
{\cal V}_k(\tau) = \frac1{\sqrt{2k_1}}\, e^{-ik_1\tau} \, e_1(k).
\end{align}
This mode corresponds to the choice of coefficients,
\begin{align}\label{norcon2}
c_{1k}=e^{\frac{i\pi}{2}\left(\nu+\frac12\right)}\,\frac{\sqrt{\pi}}{2},
\qquad d_{1k}=0.
\end{align}
Using the relations \eqref{inicon} and  \eqref{norcon2},
and adopting to the case of ${\cal U}_k$ as well,
we have the exact solutions for ${\cal V}_k$ and ${\cal U}_k$,
\begin{align}\label{exactVU}
{\cal V}_k(\tau) &= \frac{\sqrt{\pi}}{2}e^{\frac{i\pi}{2}
\left(\mu_1+\frac12\right)}
\sqrt{-\tau}\, H_{\mu_1}^{(1)}(-k_1\tau)\, e_1(k),
\nn \\
{\cal U}_k(\tau) &= \frac{\sqrt{\pi}}{2}e^{\frac{i\pi}{2}
\left(\mu_2+\frac12\right)}
\sqrt{-\tau}\, H_{\mu_2}^{(1)}(-k_2\tau)\, e_2(k).
\end{align}

As we see in the initial mode \eqref{inicon}, the effective wave number $k_1$
should be real to have a well-defined quantum state.
This means that the square of the effective wave number $k_1^2$
should be non-negative.
Therefore, the wave number $k$ should be constrained from (\ref{wavevec}) as
\begin{align} \label{stablevac}
k^2 \geq k_{{\rm min}}^2 \equiv \frac{\alpha^2 M_{\rm P}^2}{6}.
\end{align}
As we see in \eqref{VU}, the initial quantum fluctuation modes of
${\cal V}_k$ with $|k| < |k_{{\rm min}}|$ are exponentially growing
or decreasing and so they break the unitary symmetry of the quantum
system in the deep inside of the horizon.
That is, these modes are not allowed as initial quantum modes.
Therefore, there is {\it an observational lower
bound} of the comoving wavenumber in our model.
In the large scale limit $k_{1,2} |\tau| \ll 1$, the modes in \eqref{exactVU} can be
approximated as
\begin{align}\label{superV}
{\cal V}_k(\tau) &\simeq e^{\frac{i\pi}{2}\left(\mu_1+\frac12\right)}
2^{\mu_1 - \frac32} \frac{\Gamma(\mu_1)}{\Gamma(\frac32)}
\frac1{\sqrt{2k_1}}
(-k_1\tau)^{\frac12-\mu_1}\, e_1(k),
\nn \\
{\cal U}_k(\tau)&\simeq e^{\frac{i\pi}{2}\left(\mu_2+\frac12\right)}
2^{\mu_2 - \frac32} \frac{\Gamma(\mu_2)}{\Gamma(\frac32)}
\frac1{\sqrt{2k_2}}
(-k_2\tau)^{\frac12-\mu_2}\, e_2(k).
\end{align}

In the slow-roll approximation with $\xi=0$ and a small nonvanishing
value of $\alpha$, the leading contributions for the curvature
and isocurvature perturbations in \eqref{ICP} are given by
\begin{align}
{\cal R} &\simeq
{Q_\varphi\over\sqrt{2\epsilon}M_{\rm P}} + {\alpha \over 2 k \epsilon H}\dot{Q}_u
-{\alpha^2 M_{\rm P} \over 2\sqrt{2}\epsilon^{3/2}a^2H^2}Q_\varphi
\,, \\
{\cal S}&\simeq {\sqrt{2}\over 3\sqrt{\epsilon}M_{\rm P} H}\dot{Q}_\varphi
-{\alpha k\over 9 \epsilon a^2 H^2} Q_u
+{\alpha^2 M_{\rm P}\over a^2H^2}
\left(
{{Q_\varphi}\over 3\sqrt{2}\epsilon^{3/2}}
-{\sqrt{2}\over 9\epsilon^{3/2}H}\dot{Q}_\varphi
\right).
\end{align}
From these expressions, we obtain power spectrums for ${\cal R}$
and ${\cal S}$,
\begin{align}
{\cal P}_{\cal R}(k) &\equiv \frac{k^3}{2\pi^2}
\langle {\cal R}{\cal R}^*\rangle \simeq
\frac{k^3}{4\pi^2\epsilon M_{{\rm P}}^2}
\left(1- \frac{\alpha^2 M_{{\rm P}}^2}{a^2 H^2\epsilon}\right)
\langle Q_\varphi Q_\varphi^*\rangle ,
\nn \\
{\cal P}_{\cal S}(k) &\equiv \frac{k^3}{2\pi^2}
\langle {\cal S}{\cal S}^*\rangle \simeq
\frac{k^3\alpha^2}{2\pi^2\epsilon ^2}
\left[\frac{k^2}{81 a^4 H^4}\langle Q_uQ_u^*\rangle
+\frac{1}{9 a^2 H^3}
\left(\langle Q_\varphi \dot Q_\varphi^*\rangle
+\langle \dot Q_\varphi  Q_\varphi^*\rangle\right)\right].
\end{align}
Using the relations of \eqref{VUQ} and \eqref{superV},
we obtain the power spectrums
at the horizon crossing point $(k=aH)$ as
\begin{align}\label{psofRS}
{\cal P}_{{\cal R}_*}
&\simeq {H_*^2\over8\pi^2 M_{\rm P}^2} {1\over \epsilon}
\left(1+\big(2-2C\big)\eta+\big(6C-8\big)\epsilon
-{M_{\rm P}^2\over k^2}{\alpha^2\over \epsilon}\right),
\nonumber \\
{\cal P}_{{\cal S}_*} &\simeq
\biggr({H_*\over 18\pi}\biggr)^2
{\alpha^2\over \epsilon^2 k^2}
\left(1+2C-32\epsilon-18\eta - {73\alpha^2M_{\rm P}^2\over 12k^2}\right),
\end{align}
where the subscripted asterisk indicates the value at the horizon
crossing point and $C=2-\ln2 -\gamma$ with the
Euler-Mascheroni constant $\gamma \approx 0.5772$.
In the computations of power spectrums in \eqref{psofRS}, we have used
the relations,
\begin{align}
&\langle Q_\varphi Q_\varphi^*\rangle_* =
\frac1{a^2_*}\langle {\cal V}_k {\cal V}^*\rangle_*
\simeq {H_*^2\over 2 k^3} \Biggr(1+\big(6C-8\big)\epsilon
+(2-2C)\eta -{\alpha^2M_{\rm P}^2\over 12k^2}\Biggr),
\nn \\
& \langle Q_\varphi \dot Q_\varphi^*\rangle
+\langle \dot Q_\varphi  Q_\varphi^*\rangle = \frac1{a_*^3}
\left(
\langle {\cal V}_k {\cal V}_k^{'*}\rangle_*
+\langle {\cal V}_k^{'} {\cal V}_k^{*}\rangle_*
\right)
- \frac{2H_*}{a_*^2} \langle {\cal V}_k {\cal V}_k^{*}\rangle_*
\simeq {H_*^3\over k^3}\biggr(2\epsilon-\eta
-{\alpha^2M_{\rm P}^2\over 6k^2} \biggr),
\nn \\
&\langle Q_u  Q_u^*\rangle =\frac1{a_*^2}
\langle {\cal U}_k {\cal U}_k^{*}\rangle_*
\simeq {H_*^2\over 2k^3} \left(1+2\big(C-2\big)\epsilon
-{37\alpha^2M_{\rm P}^2\over 12k^2}\right).
\end{align}
The expansions of power spectrums in \eqref{psofRS}
are only valid in the range of $\alpha$,
\begin{align}\label{alpharange}
M_{{\rm P}}^2\alpha^2 \lesssim \epsilon^2 k^2.
\end{align}
In $\alpha\to 0$ limit, the power spectrums
${\cal P}_{{\cal R}}$ and ${\cal P}_{{\cal S}}$ at the horizon
crossing point in \eqref{psofRS} become those of the single field
case.
Finally, the spectral indices for ${\cal R}$ and ${\cal S}$ in the
large scale limit ($k_{1,2} |\tau| \ll 1$) are given by
\begin{align}
n_{\cal R} - 1 &\equiv \frac{d\ln {\cal P}_{\cal R}}{d\ln k}
 \simeq   2\eta-6\epsilon +
{2\alpha^2M_{\rm P}^2\over \epsilon \, k^2},
\nn \\
n_{\cal S} - 1 &\equiv \frac{d\ln {\cal P}_{\cal R}}{d\ln k}
\simeq  -2+4\eta-10\epsilon
+{67\alpha^2M_{\rm P}^2\over 6k^2}.
\end{align}
If we take (\ref{alpharange}) into account,
we find the last terms in $n_{\cal R}-1$ and $n_{\cal S} -1$
are of the order of $\epsilon$ and $\epsilon^2$, respectively.
These imply that while the curvature perturbation is  nearly scale invariant,
the isocurvature perturbation is proportional to the
inverse square of the comoving wavenumber.

The running of the scalar spectral index is given by
\begin{align}\label{running}
\frac{dn_{\cal R}}{d\ln k} &= -\frac{4\alpha^2 M_{\rm P}^2}{\epsilon k^2}
+ {\cal O}(\epsilon^2, \epsilon \eta).
\end{align}
The leading term becomes of the order of $\epsilon$ from (\ref{psofRS}) and \eqref{alpharange}.
Recent Planck~\cite{Ade:2013uln}
and WMAP nine-year~\cite{Hinshaw:2012fq} data show
\begin{align}
n_{\cal R} &= 0.9603, \quad
\frac{dn_{\cal R}}{d\ln k} = -0.0134 ~~ ({\rm Planck}),
\\
n_{\cal R} &= 0.9608,
\quad \frac{dn_{\cal R}}{d\ln k} = -0.019  ~~ ({\rm WMAP-9yr})
\end{align}
with the pivot scale $k_0 = 0.05\,{\rm Mpc^{-1}}$ (Planck) and $k_0 = 0.002
\,{\rm Mpc^{-1}}$ (WMAP), respectively. Since $\alpha^2 > 0$ and
$\epsilon >0$, the sign of the running of the spectral index
in (\ref{running}) is consistent with the observations.
However, to give some restriction on the value of $\alpha$
we need more investigations using a specific potential of the single
scalar field. We leave these issues for future work.

\subsection{Tensor mode}

Using the relations \eqref{epsH} and \eqref{ch_time}, we obtain
\begin{align}\label{appa}
\frac{a''}{a} =(aH)^2\left(2+\frac{\dot H}{H^2}\right)\simeq
{\cal H}^2\left(2- \epsilon^H\right)
\simeq \frac{2+ 3\epsilon}{\tau^2} + \frac{\alpha^2 M_{{\rm P}}^2}{6}.
\end{align}
Then the linearized perturbed equation \eqref{tmas} for the tensor mode,
up to the leading order in the slow-roll parameters and the gradient constant $\alpha$,
can be written as
\begin{align}
u_{T}^{\prime\prime} + \left[k_T^2  - \frac{\mu_T^2 -\frac14}{\tau^2}
 \right] u_{T} = 0,
\end{align}
where
\begin{align}
k_T^2 = k^2 -\frac{\alpha^2 M_{\rm P}^2}{6}
\,, \quad
\mu_T = \frac{3}{2} + \epsilon.
\end{align}
Interestingly, the tensor mode has {\it the same lower bound}
of the comoving wavenumber with that of the scalar mode in
\eqref{stablevac}.

Choosing the initial condition with a positive frequency mode,
just like \eqref{inicon} in the initial state of the scalar mode,
we obtain the following solutions in the large scale limit ($k_T |\tau|
\ll 1$),
\begin{align}
u_T &\simeq  \frac{\sqrt{\pi}}{2}e^{\frac{i\pi}{2}
\left(\mu_T+\frac12\right)}
\sqrt{-\tau}\, H_{\mu_T}^{(1)}(-k_T\tau) \, e_T(k),
\nn \\
&\simeq e^{i(\mu_T+{1\over2}){\pi\over2}} {1\over \sqrt{2 k_T}}
A(\epsilon)\big(-k_T\tau\big)^{-1}\, e_T(k),
\nn
\end{align}
where $A(\epsilon)=1-\epsilon \ln(-k \tau)
+ \epsilon \left(2-\gamma-\ln2\right)$ and $e_T(k)$
is an independent Gaussian random variable and has the same
properties with those defined in \eqref{ranGau}.

With these solutions, the power spectrum for the tensor modes $h_{ij}$
is obtained as
\begin{align}
{\cal P}_T(k) = 2 {\cal P}_h = \frac{8}{a^2 M_{{\rm P}}^2}\, {\cal P}_{u_{T}}=
{2H_*^2\over M_{\rm P}^2 \pi^2}
\left(1+\big(2C-4\big)\epsilon-{\alpha^2 M_{\rm P}^2\over 12 k^2}\right),
\end{align}
where we used $h_{\lambda} = \frac{2}{aM_{\rm P}} u_T$ and the factor ``2''
comes from the two polarization states.
We also used the quantities at the horizon crossing point ($k=aH$),
\begin{align}
&(-k_T\tau)^{n}|_* \simeq \biggr(1+\epsilon+{\alpha^2M_{\rm P}^2\over 12k^2}\biggr)^n,
\nn \\
&\langle u_T u_T^*\rangle_* \simeq {1\over 2k}
\left(
1+\big(2C-4\big)\epsilon-{\alpha^2 M_{\rm P}^2\over 12 k^2}
\right).
\end{align}
The spectral index for tensor modes and tensor-to-scalar ratio
are given by
\begin{align}
&n_{T}-1 \equiv {d \ln {\cal P}_{\cal T}\over d \ln k}\simeq -2\epsilon
-{5\alpha^2M_{\rm P}^2\over 6k^2},
\\
&r \equiv \frac{{\cal P}_T}{{\cal P}_{\cal R}} \simeq 16\epsilon_*
\left(1+(4-4C)\epsilon+(2C-2)\eta + {\alpha^2 M_{\rm P}^2\over k^2 \epsilon}
\right).
\end{align}

\section{Conclusion}

In this paper, we studied the cosmological properties of a multi-field inflation model in which we considered the single field and an additional triad of scalar fields with a non-minimal coupling which is dependent on the single scalar field.

As a special background solution, we considered spatially dependent linear profiles
for the triad of scalar fields, while the single scalar field depends on the cosmic time only.
Due to this linear property, all components of the energy momentum tensor evolve
homogeneously and isotropically on the background FRW metric.
For this reason, we can regard the background evolution
as that of the single field inflation in the presence of a matter with an
equation of state $w=-1/3$, which spreads  to the whole space.
However, there appeared one more scalar degree of freedom
from the triad of scalar fields in the perturbation level.
Therefore, our model is different form the single field inflation model
in some cosmological background.

For the exponential-type potential, we found an exact solution describing the
power-law inflation, as in the case of the single field inflation model.
We also investigated the properties of the background evolution under the
assumption of the slow-roll inflation.
The background behaviors of the slow-roll inflation are determined by the
two parameters $\xi$ and $\alpha$ which are associated with the non-minimal coupling and the gradient
constant of the triad of scalar fields, respectively.
We have drawn contour lines of the $e$-foldings for the quadratic potential on the
parameter space of $\xi$ and $\alpha$ in Fig.1.

We also investigated the properties of the scalar and tensor perturbations
for the slow-roll inflation with vanishing $\xi$ and small $\alpha$.
In the scalar fluctuations, there are two physical modes, ${\cal V}_k$
and ${\cal U}_k$ originating  from the scalar field and the triad of scalar fields
respectively in the spatially flat gauge.
The two scalar modes are decoupled from each other and have the form of
Sasaki-Mukhanov equation in the leading contribution of $\alpha$.
We found that there exists {\it an observational lower bound} of the comoving wave number,
$|k_{{\rm min}}| \sim \alpha M_{{\rm P}}$, to have
a well-defined initial quantum state for ${\cal V}_k$
with a positive energy eigenvalue.
We also obtained the power spectrums and the corresponding spectral indices
for the comoving curvature and isocurvature perturbations.
If we compare our results for the power spectrum with those of the single
field inflation model, the $\alpha$-dependent terms are always proportional
to $\alpha^2/k^2$.  Due to this reason, the spectral indices for the
curvature and isocurvature perturbations have the contribution of
$\alpha^2/k^2$ terms as well.
As  is  already well-known, the isocurvature perturbation in the single field inflation model
is vanishing in the large scale. However, the power spectrum
for the isocurvature perturbation is non-vanishing and has contributions from both
${\cal V}_k$ and ${\cal U}_k$ modes in our model.
We also implemented the same analysis for the tensor perturbation.
We found {\it the same lower bound} of the comoving wavenumber in the tensor mode.

In the case of non-vanishing $\xi$, ${\cal V}_k$ and ${\cal U}_k$ modes
are coupled with each other. Then it is difficult to obtain analytic solutions,
unlike the case of $\xi=0$. As we saw in the analysis of the background evolution,
the value of $\xi$ changes the behaviors of the cosmological evolution significantly.
Therefore, we can expect that the value $\xi$ may have some important roles
in the perturbations as well. We leave the analysis of the nonvanishing $\xi$
for future work.

One important difference between the single field model and ours is the
existence of the isocurvature perturbation in our model,
which originates from
the spatial gradient of the background solution.  One may also test
whether the nonvanishing spatial gradient can be a source of the
cosmological non-Gaussianity.
In the solid inflation model~\cite{Endlich:2012pz}, the authors
obtained unusually large non-Gaussianity, $f_{NL} \sim
1/(\epsilon c_s^2)$ with sound speed $c_s$.
Since we are considering the inflaton field in addition to the triad
of the scalar fields, we can expect that the non-Gaussianity behaviors of
our model is different from that of the solid inflation model.
It will be interesting if we compare the non-Gaussianity behavior of our model with those of the solid inflation model and the recent observational data.
The details will be reported elsewhere.

\section*{Acknowledgements}
We would like to thank Inyong Cho, Shinsuke Kawai, Eric V. Linder, Seong Chan Park,
and Uros Seljak for helpful discussions.
This research was supported by Basic Science Research Program
through the National Research Foundation of Korea (NRF) funded
by the Ministry of Education, Science and Technology (NRF-
2010-0022596) (S.K.).
This work was supported by the Korea Research Foundation Grant
funded by the Korean Government with grant numbers 2011-0009972 (O.K.),
and by the World Class University grant no. R32-10130 (O.K.).
This work was also supported by the National Research
Foundation of Korea(NRF) grant funded by the Korea government(MEST) through the Center
for Quantum Space- time(CQUeST) of Sogang University with grant number 2005-0049409 (P.O.)
and by the BSRP through the National Research Foundation of Korea funded by the MEST (2010-21
0021996) (P.O.).

\end{document}